\documentclass[nocrop]{bioinfo}

\copyrightyear{2020}
\pubyear{2020}

\application

\usepackage[hidelinks]{hyperref}
\usepackage{ifpdf}

\begin{document}
\firstpage{1}

\subtitle{Sequence Analysis}

\title[SOPanG\,2: online searching over a pan-genome without false positives]{SOPanG\,2: online searching over a pan-genome without false positives}
\author[A.~Cis{\l}ak and Sz.~Grabowski]{Aleksander Cis{\l}ak\,$^{1}$ and Szymon Grabowski\,$^{2,*}$}
\address{$^{1}$Milk Tea Analytics, Lodz, Poland,~$^{2}$Institute of Applied Computer Science, Lodz University of Technology, 
Poland}
\corresp{$^\ast$To whom correspondence should be addressed.}

\abstract{
\textbf{Motivation:}
The pan-genome can be stored as elastic-degenerate (ED) string, a recently introduced compact representation of multiple overlapping sequences. 
However, a search over the ED string does not indicate which individuals (if any) match the entire query.\\
\textbf{Results:}
We augment the ED string with sources (individuals' indexes) and propose an extension of the SOPanG (Shift-Or~for~Pan-Genome) tool to report only true positive matches, omitting those not occurring in any of the haplotypes.
The additional stage for checking the matches yields a penalty of less than 3.5\,\% relative speed in practice, 
which means that SOPanG\,2 is able to report pattern matches in a pan-genome, mapping them onto individuals, at the single-thread throughput of above 430\,MB/s on real data.\\
\textbf{Availability and implementation:} 
SOPanG\,2 can be downloaded here:~\href{https://github.com/MrAlexSee/sopang}{github.com/MrAlexSee/sopang}.\\
\textbf{Contact:}~\href{mailto://sgrabow@kis.p.lodz.pl}{sgrabow@kis.p.lodz.pl}
}

\maketitle

\section{Introduction}
Genomes within a species may vary between one another, even if slightly, and together they form what is referred to as a pan-genome. 
Using a collection of possibly thousands of haplotypes, instead of a single reference genome, reduces the bias of read aligners, inevitable if sequence reads are mapped onto a single genome.
Additionally, pan-genomic analyses can lead to a deeper understanding of the genome characteristics, evolution, acquisition of foreign genes, etc.~\citep{Medini05}.

As the level of similarity between haplotypes is 
high from the computational perspective, they are often processed together as the pan-genome rather than separately.
In particular, we use the elastic-degenerate string representation~\citep{IliopoulosKP17}, i.e.,~a sequence $T = T_0 \ldots T_{n-1}$ over an alphabet $\Sigma$, where the segment $T_i$ is a non-empty set of variable-length 
variants.
All strings within a segment are distinct and one of them, denoted by $\varepsilon$, may be of length 0.
Let us present an example of a string $T$ of {\em length} $n = 7$ with the total number of characters (called {\em size}) $N = 20$ 
($\varepsilon$ is of length 0 but its size is 1), 
which comprises four deterministic and three non-deterministic (ND)
segments:\\
$T = \{\texttt{G}\} \{\texttt{AA},\texttt{AG},\varepsilon\} \{\texttt{A}\} \{\texttt{CTG},\texttt{CAA},\texttt{AC}\} \{\texttt{A}\} \{\texttt{G},\varepsilon\} \{\texttt{CA}\}$.

Unfortunately, this simplified model admits paths not occurring in any individual.
In a former work~\citep{CislakGH18}, we presented an efficient online pattern matching algorithm for this particular representation.
In this article, however, we augment $T$ with information about so-called sources, allowing for rejecting spurious matches and mapping true matches onto individuals.
Namely, for each variant in 
each ND segment, we store indexes corresponding to individuals where the string occurs, which allows us to verify whether a given match contains variants with at least one common individual. 
Thereby, false positive matches can be eliminated. 
We retain the original elastic-degenerate representation since it allows us to perform fast processing in the initial stage, with a modest performance penalty for the later verification.
Moreover, its space requirements are smaller when compared to storing all information about the pan-genome using one of the widespread formats (in particular,~Variant Call Format).

More formally, in this work we solve the following problem, for a genome collection of $r$ individuals.
Given an elastic-degenerate sequence $T$ of length $n$ and size $N$, with $n'$ non-deterministic 
and $n''$ deterministic segments, where $n' + n'' = n$, 
a source sequence $S$ of length $n'$, where $S_i$ corresponds to $T_{i'}$, $i' \geq i$, such that $T_{i'}$ is the $i$-th non-deterministic segment in $T$, and $|S_i|$, the number of source subsets for the $i$-th non-deterministic segment is $|T_{i'}| - 1$, the number of strings in $T_{i'}$ minus one,
and given a pattern $P$ of length $m$ consisting of symbols over alphabet $\Sigma$, 
return 
sources $\mathcal{S}$, where $\forall s_j \in \mathcal{S}: s_j \in \{0, \ldots, r-1\}$
such that there exists a path in $T$ 
(possibly not covering the whole variant string in the start and in the end segment),
matching exactly the pattern $P$, 
for which the intersection of source subsets corresponding to the variants covered by the path 
contains the given source $s_j$.

Let us explain our task on an example. 
For four individuals (their corresponding indexes denoted by $0, \ldots, 3$), we might have the ED text $T = \texttt{\{AA\}\{AG,G\}\{CG,N,TT\}}$ and 
the source sequence
$S = \texttt{\{0\}\{\{0,2\}\{3\}\}}$. 
We assume that the last variant of each 
ND
segment (in the above example: \texttt{G} and \texttt{TT}, respectively) 
comes from the reference sequence. 
This indicates the following.

For the first ND segment, 
\texttt{AG} is associated with source \texttt{0}, and \texttt{G} (from the reference sequence) 
with the remaining sources \texttt{1}, \texttt{2} and \texttt{3}.
For the second ND segment, \texttt{CG} is associated with sources \texttt{0} and \texttt{2}, \texttt{N} with source \texttt{3}, and \texttt{TT} (from the reference sequence) 
with the remaining source \texttt{1}.

In every segment, each variant is associated with some individuals. 
Moreover, each individual must be associated with exactly one variant from every segment. 
The proposed human-readable format of the sources data is defined to be consistent with the ED text format.
For practical use, we employ another, compressed representation (see the following section).

The elastic-degenerate search problem has been studied under exact and approximate (Hamming and edit distance) models, both from the theoretical and the practical perspective~\citep{Bernardini20}. 
The exact search problem can be solved online with a linear dependency on $N$, notably in expected $O(nm^{1.381} + N)$ time, where $m$ is the pattern length, with the application of fast matrix multiplication~\citep{BernardiniGPPR19}. 
The bit-parallel approach~\citep{GrossiILPPRRVV17,CislakGH18,PissisR18,RautiainenMM19} offers
$O(N \lceil m/w \rceil)$ time, where $w$ is the machine word size in bits (e.g.,~64).

We also need to point out that mapping patterns (reads) onto collections of haplotypes has been traditionally attacked in the offline setting, i.e.,~with an index.
Earlier solutions, e.g.,~\citep{HuangPB13,SirenVM14}, represented both the original haplotypes and their recombinations. 
In this way they suffered from false positives occurring when a read was mapped onto an unobserved recombination, i.e.,~a substring not occurring in any of the individuals.
Some of the more recent solutions are haplotype-aware though~\citep{SirenGNPD20} and report locations of a pattern (if there are rather few of them) in microseconds, with reasonable space requirements.
Online solutions cannot get close to such search speeds for long patterns (e.g., reads of length 100), but may be more scalable for multiple pattern search and are more feasible to construct (where ``constructing'' is transforming the input 
to a convenient processing form).

\section{Materials and methods}
The proposed algorithm is based directly on our earlier work~\citep{CislakGH18}, with an additional stage for identifying true positive matches. 
The input text consists of a concatenation of deterministic and non-deterministic segments. Starting from the initial segment, we process the segments one-by-one. 
If the current segment is deterministic, we run the standard Shift-Or procedure~\citep{BYG92} over it.
Otherwise, for a non-deterministic segment with $u$ strings in it, we create $u$ copies of the state vector and update each of them with the corresponding string. 
After processing all $u$ strings in the segment, the $u$ bit-vectors are AND'ed. 
The unset bits in the result correspond to active states (i.e.,~matching pattern prefixes) over any path ending in some variant in the current segment. 
The algorithm continues in this fashion until all segments have been processed.

After obtaining the initial matches, which contain both true positives and false positives (where false positives correspond to pan-genome paths not occurring in any individual), we proceed to the final stage. 
Conceptually, for a given match, we build a tree of all possible paths through the elastic-degenerate string that might correspond to this match. 
We start in the segment where the match ended and move towards the beginning of the string. 
Each tree level corresponds to a single segment and each node corresponds to a single variant.

A node stores a position of the matching variant in the pattern. 
A node is connected with a child, i.e.,~a variant from the previous segment, only if ($i$) pattern substrings match both variants at the respective positions, and ($ii$) the two variants share at least one common source.

We proceed in this manner until all letters from the pattern have been exhausted. 
A match is verified as a true positive if there exists a path from the root (the variant of a segment where the match ended) towards one of the leaves. 
In the actual implementation, the tree is 
maintained as a flat data structure storing all nodes at the currently processed level,
and checking for sources shared between the variants is done with the use of a bitset.

As mentioned previously, all source information is stored in a custom variable-length differential encoding format. 
Each segment start is marked with a byte flag, followed by a 
number indicating the count of individuals in this variant. 
Later, we store each source index in an increasing order and as a difference between this index and the previous one (with the obvious exception of the first index which is stored as is). 
Moreover, each number occupies either one or two bytes, using a simple variable-length scheme, limiting the number of individuals to 16,383.
At the end, the entire dataset is compressed with the zstd library (\href{https://facebook.github.io/zstd}{facebook.github.io/zstd}).

Our algorithm is intended to be practical, without worst-case guarantees, yet let us take a brief look on its average case.
Let $occ'$, $occ' \leq N$, be the number of end match positions for pattern $P$.
For a given end match position, which belongs to a segment $T_i$, we first find all variants in $T_i$ whose prefixes ending at that position match the suffix of $P$, and perform the bitwise OR over their source-related bitsets.
Assuming $r$ individuals and $O(1)$ number of variants in a segment, this takes $O(\lceil r/w \rceil)$ time.
Let us now assume that in every segment no variant is a (proper) suffix of another.
This means that at the beginning of segment $T_{i-1}$ only one variant can be ``active'', i.e.,~matching a suffix of $P$ (when followed with a prefix of $T_i$).
We find this variant (if any) in time (at most) linear in the number of symbols in segment $T_{i-1}$ plus $O(\lceil r/w \rceil)$ time for bitwise operations.
We continue in the same manner with the preceding segments, until we find the start segment, 
in which (by analogy to the end segment, $T_i$) there can be more than one variant with a suffix matching the prefix of pattern $P$.
This again is handled in $O(\lceil r/w \rceil)$ time with our assumptions.
In sum, with $N/n$ symbols in a segment on average, we spend (at most) $O(m (N/n + \lceil r/w \rceil))$ average time for candidate verification, which gives 
$O(occ' \times m (N/n + \lceil r/w \rceil))$ overall time.

In theory, we do not have to read naively all variants in a segment, but can build a trie for each segment and follow it during match extensions.
For a constant-size alphabet, a trie is built in linear time, $O(N)$ in total.
This change, however, does not improve our average-case time complexity, provided that the number of variants in a segment is constant, and is also unlikely to be practical.

The most important assumption we made in our analysis is that the set of variants in every segment is suffix-free. 
Unfortunately, this is not the case for real data, however, otherwise the analysis seems much more complex.
Nonetheless, the experimental results show that in practice handling all candidates is relatively fast.

\section{Results}
SOPanG\,2 is implemented in C++ (compiled using 64-bit gcc~7.4.0 with \texttt{-O3}~flag) and tested on 64-bit Ubuntu Linux 17.10 (Intel i7-4930K@3.4\,GHz CPU, 64\,GB DDR3 RAM).
Experiments were run on both real and synthetic data.
The former were the human genomic data from Phase 3 of 1000\,GP.

The input was given as a file pair: an elastic-degenerate string and sources in the aforementioned custom format. 
For computational purposes, we treat diploids of each individual as two distinct individuals, obtaining 5,008 haploid sequences in total.
Both files were generated from the FASTA reference sequence and variations data in VCF format.

The synthetic data had 100\,k, 500\,k, 1\,M and 1.6\,M positions (in the sense where each deterministic segment contains a single letter)~\citep{GrossiILPPRRVV17}.
In each case, the number of non-deterministic positions was equal to 10\,\%, the maximum number of variants in each non-deterministic segment was equal to 10, and the maximum length of each variant string was equal to 10. 
Moreover, for synthetic data we have tested the following individual counts: 16, 128, 1,024, and 8,192.

We offer two modes for matching with sources: a true-false check if there is at least one individual corresponding to the match, and full matching where for each segment index the complete list containing individuals' indexes is returned.
The former is slightly faster, since we can report the result as soon as a single true positive is found.
Note that in both procedures we need to store the variant index as well as the exact position in the variant where the match ended. 
This is in addition to the segment index where the match ended, which was sufficient for regular ED text searching.

\begin{figure}
\centerline{
\includegraphics[scale=0.16]{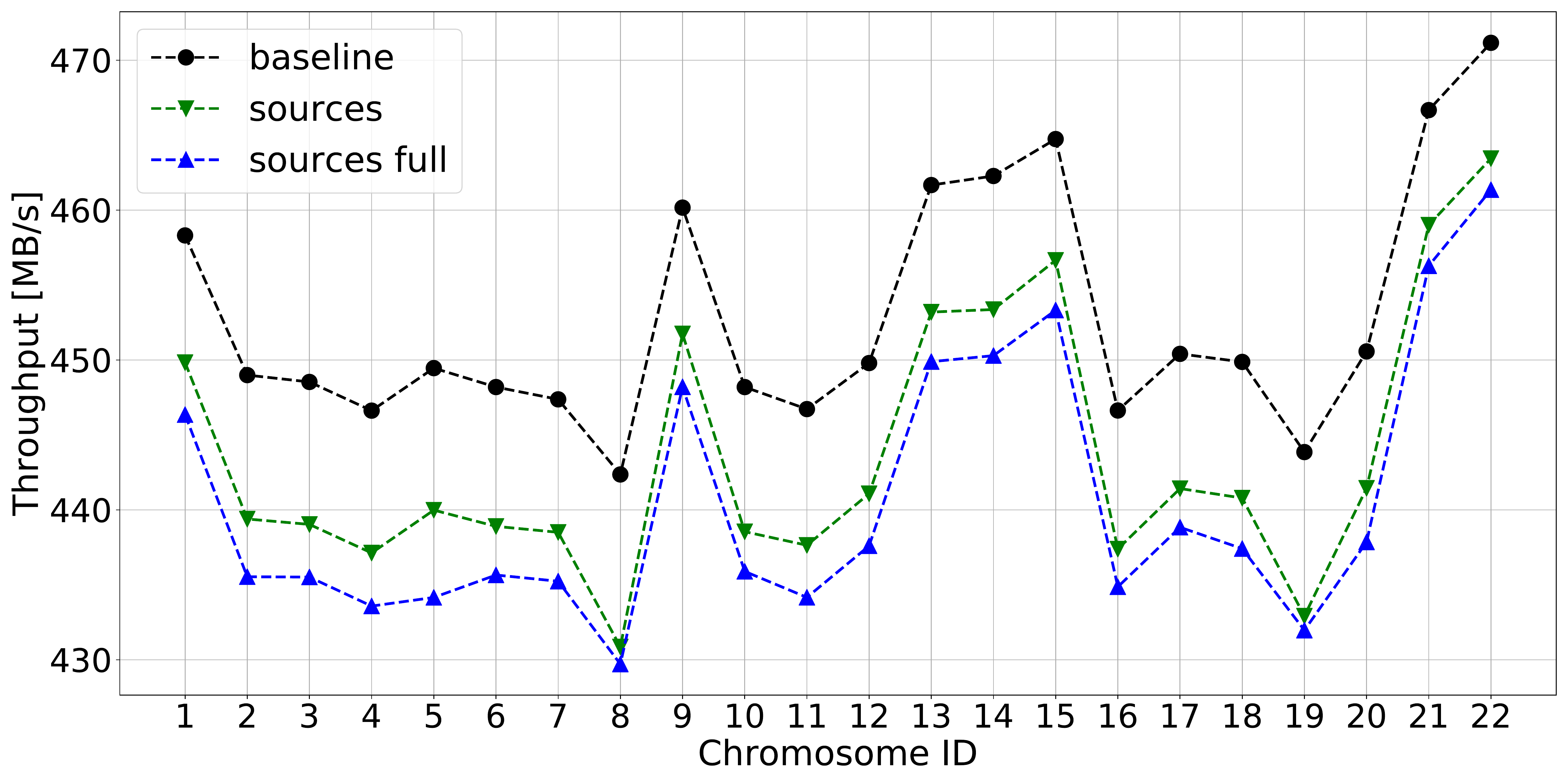}
}
\caption[Illustration]
{SOPanG\,2 throughput on real data (1\,MB = $10^6$\,B) for $m=8$.
The line ``baseline'' reveals the performance of finding matches in the pan-genome only, ``sources'' answers whether a given match is genuine (at least one individual contains the traversed path), and ``sources~full'' reports the complete list of matching individuals.}
\label{fig:sopang2}
\end{figure}

In practice, performance difference between discussed methods is not significant.
Figure~\ref{fig:sopang2} shows that our solution achieves search throughput above 430\,MB/s (where 1\,MB = $10^6$\,B, measured with respect to $N$) on the real-world data, for chromosomes labeled 1 to 22. Relative slowdown oscillated in the range 1.6--2.6\,\% for verification matching and 2.0--3.5\,\% for full matching with sources. 
This refers to measured CPU time after the data has been loaded into memory, decompressed, and parsed.
Our search algorithm is insensitive to the pattern length for the tested set of 8, 16, 32, and 64 characters, and the slowdown rate for the verification depends directly on the number of matches to be verified (which is generally higher for shorter patterns).
The baseline throughput is higher than reported previously~\citep{CislakGH18}, thanks to minor implementation improvements (the hardware setup remains unchanged).

For synthetic data, the difference between matching with and without sources was not observed. 
For four synthetic datasets having sizes described above, the throughput was equal approximately to 313, 305, 304, and 304\,MB/s, respectively. 
A smaller slowdown than the one reported for real-world data was probably caused by the overall lower number of matches for the tested patterns.

Let us also indicate one limitation of the use of sources which is related to increased working memory usage.
It stems from the need to keep the sources map for the whole chromosome in memory and results in memory consumption spike up to the order of magnitude (e.g.,~for chromosome 12, for the entire executable, with max individual count limited at compilation time, approximately 10.8\,GB for sources vs 1.5\,GB for regular matching).

Part of the original motivation behind the introduction of the elastic-degenerate string format was to represent the data in a compact form. 
Let us consider disk space requirements for the 22 tested chromosomes. All pairs comprising an elastic-degenerate text and a corresponding sources file in our compressed format measure around 2.6\,GB in total.
On the other hand, the raw data in FASTA and VCF formats contain a lot of information that is redundant for 
our purposes.
The combined total size for such data 
(compressed with gzip, as available on the 1000\,GP website) exceeds
15\,GB, measuring roughly 5.8 times more.

\section{Conclusions}
We presented an uncomplicated and efficient extension of our previous algorithm for exact pattern matching over a pan-genome.
While still using an elastic-degenerate text format, we eliminated its inherent limitation of false positive matches (i.e.,~matches not present in any individual genome) by augmenting it with the information about segment variant individuals.
In our implementation, the search speed remains almost unaffected in practice.
The data size increases roughly fourfold, but it is still much smaller than for the full information in FASTA and VCF formats.
For a single-pattern online search where the pattern size does not exceed the machine word size, we still maintain 
an order of magnitude speedup over
available implementations~\citep{GrossiILPPRRVV17, PissisR18}.

In our previous work~\citep{CislakGH18}, we presented preliminary results regarding approximate bit-parallel matching for the elastic-degenerate format. 
We believe that in the future that technique could be also used for the model with sources.
Moreover, different formats for storing the sources data could be defined, e.g.,~interleaving the elastic-degenerate text and sources in a single file for disk-I/O-aware algorithms.

Furthermore, a block-based ``semi-indexing'' approach~\citep{GrabowskiSR17,BingmannBGI19} can probably be used for our scenario, where most of the pan-genome blocks are quickly eliminated as they do not contain all the $k$-mers from the query pattern.

\end{document}